\documentclass[aps,prl,twocolumn,showpacs,preprintnumbers,superscriptaddress,amsmath,amssymb]{revtex4}
\usepackage{graphicx}
\usepackage{dcolumn}
\usepackage{bm}

\begin{document}

\title{Evidence of $s$-wave pairing symmetry in layered superconductor
Li$_{0.68}$NbO$_2$ from the specific heat measurement}

\author{G. T. liu}
\author{J. L. Luo}
\email{JLLuo@aphy.iphy.ac.cn}
\affiliation{Beijing National
Laboratory for Condensed Matter Physics, Institute of Physics,
Chinese Academy of Sciences, Beijing 100080, People's Republic of
China}

\author{Z. Li}
\affiliation{Beijing National Laboratory for Condensed Matter
Physics, Institute of Physics, Chinese Academy of Sciences,
Beijing 100080, People's Republic of China}

\author{Y. Q. Guo}
\affiliation{Beijing National Laboratory for Condensed Matter
Physics, Institute of Physics, Chinese Academy of Sciences,
Beijing 100080, People's Republic of China}

\author{N. L. Wang}
\affiliation{Beijing National Laboratory for Condensed Matter
Physics, Institute of Physics, Chinese Academy of Sciences,
Beijing 100080, People's Republic of China}

\author{D. Jin}
\affiliation{Beijing National Laboratory for Condensed Matter
Physics, Institute of Physics, Chinese Academy of Sciences,
Beijing 100080, People's Republic of China}

\author{T. Xiang}
\affiliation{Institute of Theoretical Physics and
Interdisciplinary Center of Theoretical Studies, Chinese Academy
of Science, Beijing 100080, People's Republic of China}

\date{\today}

\begin{abstract}
A high quality superconducting Li$_{0.68}$NbO$_2$ polycrystalline
sample was synthesized by deintercalation of Li ions from
Li$_{0.93}$NbO$_2$. The field dependent resistivity and specific
heat were measured down to 0.5 K. The upper critical field $H_{c2}
(T)$ is deduced from the resistivity data and $H_{c2}(0)$ is
estimated to be $\sim 2.98$ T. A notable specific heat jump is
observed at the superconducting transition temperature $T_c \sim
5.0$ K at zero field. Below $T_c$, the electronic specific heat
shows a thermal activated behavior and agrees well with the
theoretical result of the BCS $s$-wave superconductors. It
indicates that the superconducting pairing in Li$_{0.68}$NbO$_2$
has $s$-wave symmetry.
\end{abstract}

\maketitle

The recent discovery of superconductivity in
Na$_{0.3}$CoO$_2$$\cdot$1.3H$_2$O\cite{Takada} has stimulated
great interest in the investigation of physical properties of
quasi-two-dimensional materials with frustrations. Among them,
lithium niobium oxide, LiNbO$_2$, has received considerable
attention since superconductivity was discovered upon partial
deintercalation of lithium atoms in this
material\cite{Geselbracht}. This compound is of certain
characteristics of high-$T_c$ cuprates. There exists a strong
hybridization of Nb 4$d$ states with O 2$p$ states, and an
elevated density of oxygen states at the $E_F$\cite{Kellerman,
Moshopoulou}. Since many layered oxides, including high-$T_c$
cuprates, Sr$_2$RuO$_4$, and Na$_{0.3}$CoO$_2$$\cdot$1.3H$_2$O,
have shown unconventional superconductivity, it is of fundamental
interest to know whether the superconducting pairing of this
material is conventional or unconventional.

In this paper, we report the experimental data of the field
dependent resistivity and specific heat measurements of a high
quality Li$_{0.68}$NbO$_2$ sample. From the data, we obtain the
temperature dependence of the upper critical field and the
electronic specific heat by subtracting the phonon contribution
from the total specific heat. By comparison with the theoretical
curves for the BCS $s$- and $d$-wave superconductors, we find that
the electronic specific heat agrees very well with the $s$-wave
curve. The low temperature electronic specific heat shows a
thermally activated behavior. It decays exponentially with
decreasing temperature. Our results indicate that
Li$_{0.68}$NbO$_2$ is a weak-coupling BCS $s$-wave superconductor.

As shown in Fig. 1, Li$_x$NbO$_2$ has a layered structure
analogous to MoS$_2$. Along the $c$-axis, the lithium planes and
NbO$_6$ trigonal-prismatic layers are stacked
alternatively\cite{Meyer}. In each Nb-O layer, Nb atoms form a
triangular lattice, similar to that of Co atoms in Na$_x$CoO$_2$.
The lithium ions occupy octahedral holes between
trigonal-prismatic Nb-O layers. Under the influence of the
trigonal crystal field, the Nb 4$d$ energy levels split into a
pattern with $d_{z^2}<d_{x^2-y^2}, d_{xy}<d_{xz},
d_{yz}$\cite{Burdett}. In the stoichiometric compound LiNbO$_2$,
the $d_{z^2}$ band is completely occupied, the unoccupied part of
conducting band is well separated from the lower one by 1.5
eV\cite{Burdett, Novikov, Turzhevsky, Ylvisaker}. So it exhibits
semiconducting behavior in low temperatures. However, in
Li$_x$NbO$_2$ with partial removal of lithium, the conduction band
that contain holes is formed\cite{Novikov}. The metallic behavior
is observed and superconductivity occurs with T$_c$ of $\sim$ 5 K.

\begin{figure}
\includegraphics[width=0.7\linewidth]{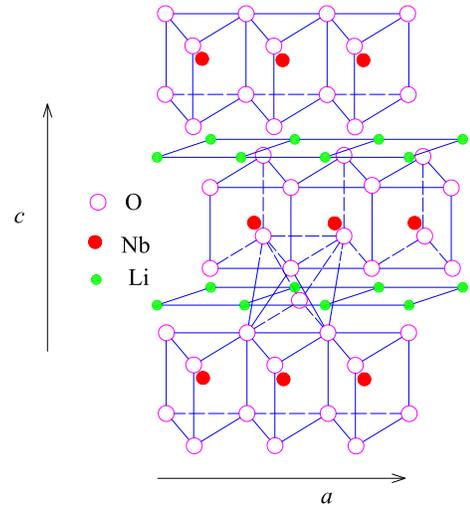}
\caption{\label{fig:fig1} {Schematic crystal structure of
LiNbO$_2$ (from Ref. \onlinecite{Geselbracht}). The lithium ions
(small open circles) occupy the octahedral holes between
trigonal-prismatic Nb-O layers. The niobium (solid circles) and
oxygen ions (large open circles) form NbO$_6$ trigonal prisms.}}
\end{figure}

The burgundy-red polycrystalline LiNbO$_2$ was prepared by heating
Li$_3$NbO$_4$ and NbO in a molar ratio of 1:2 in an evacuated
fused-silica tube at 1050$^{\circ}$C for 60 h. Pressed pellets of
the reaction mixture were wrapped in niobium sheet to avoid
reacting with the quartz. Li$_3$NbO$_4$ was obtained by heating a
mixture of Li$_2$CO$_3$ and Nb$_2$O$_5$ in a molar ratio of 3:1 at
900$^{\circ}$C in air for 48 h, and NbO was prepared by firing a
mixture of Nb$_2$O$_5$ and metallic Nb in a molar ratio 1:3 in an
evacuated fused-silica tube at 1100$^{\circ}$C for 72 h. The
lithium content was determined by inductively coupled plasma
spectroscopy. The as-prepared sample was slightly
non-stoichiometric in lithium due to its volatility,
Li$_{0.93}$NbO$_2$, which is the same as previous
reports\cite{Geselbracht, Kumada, Geselbracht2}.
Li$_{0.68}$NbO$_2$ was prepared from Li$_{0.93}$NbO$_2$ by
chemical treatments at room temperature using bromine or
hydrochloric acid as oxidative reagent. After the treatment, the
sample changed from reddish to black, indicating that the
electronic properties were modified. Powder x-ray diffraction
pattern obtained by a M18AHF x-ray diffractometer using
Cu$K_{\alpha}$ radiation showed that both the as-prepared and
deintercalated samples are single phased with no trace of
LiNbO$_3$ and Nb.

The low temperature resistivity was determined by the standard
four point measurement with a Quantum Design PPMS. The low
temperature specific heat measurements in the range from 0.5 to 7
K were performed with a $^3$He heat-pulsed thermal relaxation
calorimeter attached to the PPMS up to 3 Tesla. The precision of
the measurement is about 1\%. The field dependence of thermometer
and addenda was carefully calibrated before the specific heat was
measured.

\begin{figure}
\includegraphics[width=0.9\linewidth]{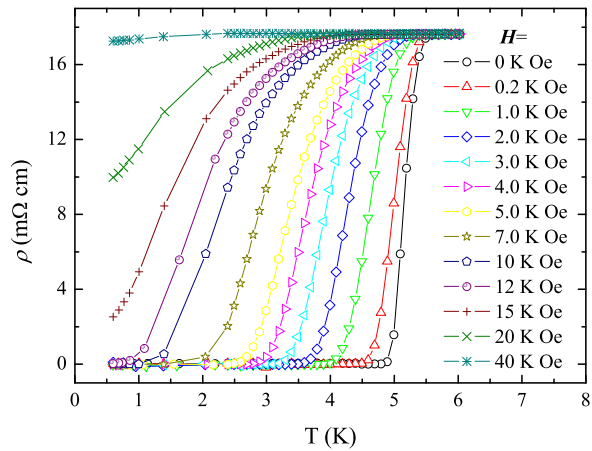}
\caption{\label{fig:fig2} {Temperature dependence of the
resistivity $\rho (T)$ for Li$_{0.68}$NbO$_2$ at different
magnetic fields.}}
\end{figure}

Figure 2 shows the temperature and field dependence of the
resistivity for Li$_{0.68}$NbO$_2$. At zero field, a
superconducting transition begins at 5.5 K, and the resistivity
becomes zero at about 4.8 K. The transition width is about 0.7 K.

In Fig. 3, the temperature dependence of the upper critical field
$H_{c2}(T)$, determined from the temperature at which the
resistivity is equal to 90\% of its normal state value for a given
field, is shown. From the Ginzburg-Laudau theory\cite{Tinkham}, it
is known that $H_{c2}$ is inversely proportional to the square of
the coherence length $\xi$:
\begin{eqnarray}
H_{c2} & = & \frac{\Phi_0}{2\pi\xi^2},\\
\xi^2 & = & \frac{\hbar^2}{2m^\ast \alpha (T)},
\end{eqnarray}
where $\Phi_0$ is the flux quanta,
$\alpha(T)\propto(1-t^2)/(1+t^2)$ and $t$=$T/T_c$ is the reduced
temperature. Therefore, $H_{c2}(T)$ can be expressed as:
\begin{equation}
{H_{c2}(T)}={H_{c2}(0)\frac{1-t^2}{1+t^2}}. \label{Hc2}
\end{equation}
By fitting the experimental data with the above equation (solid
curve in Fig. 3), we find that the zero temperature upper critical
field $H_{c2}(0)\approx 2.98$ T. The coherence length
$\xi_{\textrm{GL}}(0)$ estimated from this value of $H_{c2}(0)$ is
$\sim$ 106 $\textrm{\AA}$, which is very close to the coherence
length of Na$_{0.3}$CoO$_2$$\cdot$1.3H$_2$O (100
$\textrm{\AA}$)\cite{Chou}.

\begin{figure}
\includegraphics[width=0.9\linewidth]{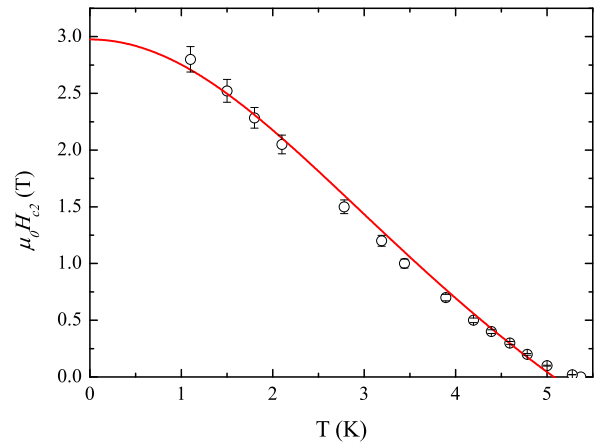}
\caption{\label{fig:fig3} {The upper critical field as a function
of temperature. The solid line is fitting curve with Eq.
(\ref{Hc2}). }}
\end{figure}

The specific heat measurement is a powerful tool for investigating
the low-lying superconducting quasiparticles. It probes bulk
properties and can be used to determine the pairing symmetry of
superconductors\cite{Moler, Chen, Revaz, Wright, Luo1}. Figure 4
shows the measurement data of the specific heat for
Li$_{0.68}$NbO$_2$. A clear anomaly associated with the
superconducting transition is discerned at about $T\sim$ 5 K at
zero field. This anomaly persists with $H$ up to 3 T. There is no
upturn in all the curves of $C/T$ down to 0.5 K. This suggests
that the sample is free of magnetic impurities.

The measured specific heat $C$ contains the contribution from both
electrons and phonons, $C = C_{el} + C_{ph}$. In low temperatures,
the phonon contribution to the specific heat $C_{ph}$ does not
depend on the applied magnetic field and follows the Debye law,
i.e. $C_{ph} \sim T^3$, in both the normal and superconducting
states. In the normal state, the electronic contribution to the
specific heat $C_{el}$ varies linearly with $T$, thus the total
specific heat can be expressed as
\begin{equation}
{C_n(T)}=\gamma_nT + \alpha T^3. \label{Cn}
\end{equation}
where $\gamma_n$ and $\alpha$ are temperature independent
coefficients. By fitting the normal state data with this formula,
we find that $\gamma_n =3.588 \pm 0.038 \,\textrm{mJ}/\textrm{mol}
\, \textrm{K}^2$ and $\alpha = 0.07258 \pm 0.00075\,
\textrm{mJ}/\textrm{mol}\, \textrm{K}^4$. This $\gamma_n$ value is
smaller than the corresponding value for Li$_x$NbS$_2$ ($\sim
\,10\, \textrm{mJ}/\textrm{mol} \, \textrm{K}^2$) \cite{Dahn},
NaCoO$_2$ ($\sim \, 24 \, \textrm{mJ}/\textrm{mol} \,
\textrm{K}^2$) \cite{Jin, Luo}, LiTi$_2$O$_4$ ($\sim\, 19.15 \,
\textrm{mJ}/\textrm{mol} \, \textrm{K}^2$)\cite{Sun} and
Sr$_2$RuO$_4$ ($\sim\, 40 \, \textrm{mJ}/\textrm{mol} \,
\textrm{K}^2$) \cite{Nishizaki}. The Debye temperature deduced
from $\alpha$ is $462\pm 12 \textrm{K}$.

\begin{figure}
\includegraphics[width=0.80\linewidth]{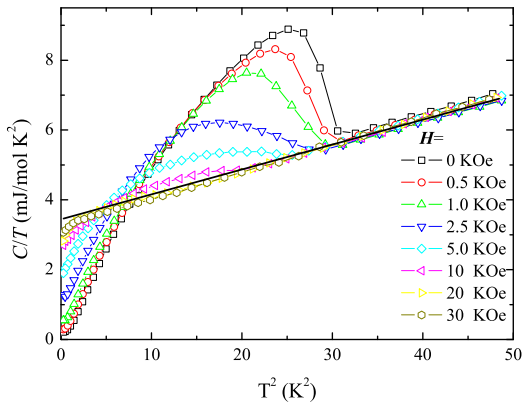}
\caption{\label{fig:fig4} {$C/T$ versus $T^2$ at various applied
fields for Li$_{0.68}$NbO$_2$. The solid line is a linear fit to
the normal state specific heat.}}
\end{figure}

The normal state electronic specific heat coefficient $\gamma_n$
is proportional to the density of states of electrons at the Fermi
level $N(E_F)$. In the free electron gas model, it can be
expressed,
\begin{equation}
{\gamma_n}=\frac{\pi^{2}}{3}k_{B}^{2} N(E_F). \label{gamma_n}
\end{equation}

Novikov et. al calculated band structure of Li$_{0.5}$NbO$_2$
using the full potential linear-muffin-tin-orbital(FLMTO)
method.\cite{Novikov} They found the density of states $N(E_F)$ at
Fermi level is 6.02 1/eV. If we take this value for
Li$_{0.68}$NbO$_2$, the value of calculated $\gamma_n$ $\sim$ 3.85
$\textrm{mJ}/\textrm{mol} \, \textrm{K}^2$ is close to
3.58$\textrm{mJ}/\textrm{mol} \, \textrm{K}^2$, the value obtained
from the specific heat measurement. This signify that the
electron-phonon coupling is very weak in this material.

In the superconducting state, the electronic specific heat can be
obtained by subtracting the phonon term from the total specific
heat $C_{el} = C - C_{ph}$. As shown in Fig. 5(a), $C_{el}/T$
drops with decreasing temperature. In the absence of magnetic
field, $C/T$ extrapolates to a small but finite value $\gamma_s =
0.195 \, \textrm{mJ} / \textrm{mol}\, \textrm{K}^2$ at zero
temperature. This residual specific heat indicates that there is a
residual density of states at the Fermi level, which in turn means
that the superconducting volume fraction of the sample is less
than 100\%. From the ratio $(\gamma_n - \gamma_s)/\gamma_n$, we
estimate the superconducting volume fraction of the sample to be
94.5\%. This value of the superconducting volume fraction is
rather high compared with the previous reports\cite{Geselbracht,
Tyutyunnik}. It is presumably due to the improved treatment of the
sample quality.

\begin{figure}
\includegraphics[width=0.90\linewidth]{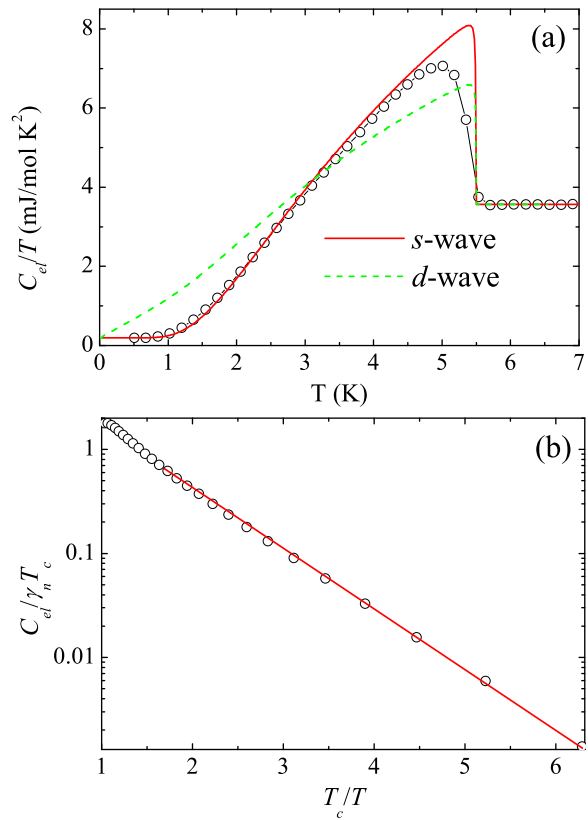}
\caption{\label{fig:fig5} {(a) $C_{el}/T$ as a function of
temperature for Li$_{0.68}$NbO$_2$. The solid and dashed lines are
the theoretical curves for the BCS $s$- and $d$-wave
superconductors, respectively. (b) $C_{el}/\gamma_nT_c$ versus
$T_c/T$. The solid line is an exponential fit to the low
temperature data after subtracting the residual specific heat
contributed from the normal volume fraction.}}
\end{figure}

Figure 5(a) compares the measurement data with the BCS mean-field
results of $s$- and $d$-wave superconductors. Apparently, the
$d$-wave result deviate significantly from the experimental data
in the whole temperature range. However, the $s$-wave result fits
extremely well with the experimental data, especially in the low
temperature regime. The jump of the specific heat at $T_c$ is less
than the theoretical value. This is probably due to the fact that
the superconducting transition of this material happens in a
finite temperature range ($0.7 \,$ K) rather than just at one
point as in the BCS theory. By plotting $C_{el}$ as a function of
$T_c/T$ in a semi-logarithmic scale (Figure 5(b)), we find that
$C_{el}$ indeed decreases exponentially with $T$ in low
temperatures. By further fitting the low temperature data with the
formula\cite{Ketterson}
\begin{equation}
C_{el} \approx 2\sqrt{2\pi}k_B N(E_F) \Delta(0)
\left(\frac{\Delta(0)}{k_BT} \right)^{3/2} e^{-\Delta(0) / k_B T},
\end{equation}
we find that $\Delta(0) \approx 0.8\, \textrm{meV}$. This value of
$\Delta(0)$ is close to that determined simply from $T_c$ using
the weak coupling BCS formula $\Delta(0) = 1.76 k_BT_c  = 0.76 \,
\textrm{meV}$.

In conclusion, we have measured the field and temperature
dependence of resistivity and specific heat of Li$_{0.68}$NbO$_2$.
From the resistivity data, the upper critical field is deduced and
analyzed using the Ginzberg-Landau theory. A notable specific heat
jump is observed at $T_c \sim 5.0 \,$ K at zero field and
suppressed by the applied field. By subtracting the phonon
contribution, the electronic specific heat $C_{el}$ is obtained
from the measured data. The linear coefficient of $C_{el}$ is
found to be $\gamma_n = 3.588\,\textrm{mJ}/\textrm{mol \, K}^2$
and the Debye temperature is $\Theta_D = 462 \,$ K. Below $T_c$,
$C_{el}$ shows a thermal activated behavior and the temperature
dependence of $C_{el}$ agrees well with the BCS result for a
$s$-wave superconductor.

We would like to thank G. C. Che for useful discussions and  H.
Chen for the help in the XRD experiments. This work was supported
by the National Natural Science Foundation of China, the Knowledge
Innovation Project of Chinese Academy of Sciences, and National
Basic Research Program of China under Contracts No. 2003CCC01000
and No. 2005CB32170X.

\end{document}